\documentclass[preprint2]{aastex631}

\usepackage{graphicx}
\usepackage{natbib}
\usepackage{booktabs}
\usepackage{amsmath, physics}
\usepackage{enumitem}

\newcommand{\Cornell}{Cornell Center for Astrophysics and Planetary Science, and Department of Astronomy, Cornell University, Ithaca, NY 14853, USA}
\newcommand{\BL}{Breakthrough Listen, University of California, Berkeley, CA 94720, USA}
\newcommand{\UCB}{Department of Astronomy, University of California, Berkeley, CA 94720, USA}
\newcommand{\Caltech}{Cahill Center for Astronomy and Astrophysics, MC 249-17, California Institute of Technology, Pasadena, CA 91125, USA}
\newcommand{\SETI}{SETI Institute, 339 N Bernardo Ave Suite 200, Mountain View, CA 94043, USA}
\newcommand{\NIJ}{Department of Astrophysics/IMAPP, Radboud University, Nijmegen, The Netherlands}
\newcommand{\KZA}{University of Malta, Institute of Space Sciences and Astronomy, Malta}
\newcommand{\Curtin}{International Centre for Radio Astronomy Research, Curtin University, WA 6102, Australia}

\shorttitle{Periodic extraterrestrial beacon searches}
\shortauthors{Suresh et al.}

\begin{document}

\title{A 4--8~GHz Galactic Center Search for Periodic Technosignatures}

\author[0000-0002-5389-7806]{Akshay Suresh}
\affiliation{\Cornell}
\affiliation{\BL}

\author[0000-0002-8604-106X]{Vishal Gajjar}
\affiliation{\BL}
\affiliation{\SETI}

\author[0000-0002-1386-0603]{Pranav Nagarajan}
\affiliation{\BL}
\affiliation{\UCB}
\affiliation{\Caltech}

\author[0000-0001-7057-4999]{Sofia Z. Sheikh}
\affiliation{\SETI}
\affiliation{\BL}

\author[0000-0003-2828-7720]{Andrew P. V. Siemion}
\affiliation{\BL}
\affiliation{\SETI}
\affiliation{\NIJ}
\affiliation{\KZA}

\author[0000-0002-7042-7566]{Matt Lebofsky}
\affiliation{\BL}

\author[0000-0001-6950-5072]{David H. E. MacMahon}
\affiliation{\BL}

\author[0000-0003-2783-1608]{Danny C. Price}
\affiliation{\Curtin}
\affiliation{\BL}

\author[0000-0003-4823-129X]{Steve Croft}
\affiliation{\UCB}
\affiliation{\BL}
\affiliation{\SETI}

\begin{abstract}
Radio searches for extraterrestrial intelligence have mainly targeted the discovery of narrowband continuous-wave beacons and artificially dispersed broadband bursts. Periodic pulse trains, in comparison to the above technosignature morphologies, offer an energetically efficient means of interstellar transmission. A rotating beacon at the Galactic Center (GC), in particular, would be highly advantageous for galaxy-wide communications. Here, we present {\tt blipss}, a CPU-based open-source software that uses a fast folding algorithm (FFA) to uncover channel-wide periodic signals in radio dynamic spectra. Running {\tt blipss} on 4.5~hours of 4--8~GHz data gathered with the Robert C.~Byrd Green Bank Telescope, we searched the central $6\arcmin$ of our Galaxy for kHz-wide signals with periods between 11--100~s and duty cycles ($\delta$) between 10--50$\%$. Our searches, to our knowledge, constitute the first FFA exploration for periodic alien technosignatures. We report a non-detection of channel-wide periodic signals in our data. Thus, we constrain the abundance of 4--8~GHz extraterrestrial transmitters of kHz-wide periodic pulsed signals to fewer than one in about 600,000 stars at the GC above a 7$\sigma$ equivalent isotropic radiated power of $\approx 2 \times 10^{18}$~W at $\delta \simeq 10\%$. From an astrophysics standpoint, {\tt blipss}, with its utilization of a per-channel FFA, can enable the discovery of signals with exotic radio frequency sweeps departing from the standard cold plasma dispersion law.
\end{abstract}

\section{Introduction} \label{sec:intro}
The Search for Extraterrestrial Intelligence (SETI, review: \citealt{Tarter2001}) is an active quest to find evidence of advanced alien life in the universe through signatures of their technologies. Radio SETI has been ongoing since the early 1960s \citep{Drake1960}. Two broad categories of potential radio extraterrestrial intelligence (ETI) signals are intentional beacon emissions \citep{Cocconi1959,Drake1973,Tarter1980,Valdes1986,Steffes1994,Mauersberger1996} and leakage radiation emanating from alien technologies \citep{Guillochon2015,Benford2016}. Of these two technosignatures, the spectrotemporal characteristics of the latter are much harder to speculate. In addition, such leakage signals are likely to be weaker. Hence, modern radio SETI efforts have primarily focused on wideband searches for deliberate narrowband ($\Delta \nu \sim 1$~Hz) Doppler-drifting beacons from Galactic planetary systems \citep{Siemion2013,Harp2016,Tingay2016,Enriquez2017,Gray2017,Margot2018,Margot2021,Pinchuk2019,Price2020,Gajjar2021,Lacki2021,Traas2021,Franz2022,Garrett2023,Tusay2022} and neighboring galaxies \citep{Gray2017,Lacki2021,Garrett2023}. \\

Broadband transient pulses enable signal detection agnostic to the choice of receiver center frequency \citep{Clancy1980}. While such pulses experience dispersion and scattering from astrophysical plasma along the line of sight, ETI might introduce negative dispersion as a means of adding artificiality \citep{Demorest2004,Siemion2013}. For example, broadband radio pulses propagating through the interstellar medium undergo cold plasma dispersion, resulting in  chromatic pulse arrival times, $t_{\rm a} \propto \nu^{-2}$. By introducing negative dispersion, ETI can intentionally overcorrect for astrophysical dispersion and transmit artificial bursts with delayed pulse arrival at high radio frequencies. Alternatively, artificiality can be incorporated through a repeating non-physical burst sequence such as a Fibonacci series (e.g., \citealt{Clement2022}). \\

Allowing for both natural and artificial dispersion, \citet{Gajjar2022} conducted extensive searches for broadband pulsed beacons from 1883 stars in the Milky Way. Their  investigations reveal that $\lesssim 1$ in 1000~stars transmit pulses repeating $\leq 500$~s with equivalent isotropic radiated power (EIRP) $\gtrsim 10^{22}$~W  at 4--8~GHz. Moreover, the power budget for transmission can be significantly lowered by incorporating periodicity in pulse emission. \\

Periodic pulsed beacons offer an inexpensive means of communication across vast interstellar distances. \citet{Sullivan1991} proposed a set of plausible pan-Galactic SETI periods in the $10^{-6}$--$10^5$~s range, some of which are based on known pulsar periods in our galaxy. Further, \citet{Edmondson2003} suggested that ETI might potentially broadcast signals mimicking millisecond pulsar periodicities for interstellar communication. Transmission of such signals from their respective anti-pulsar directions relative to the Earth would then indicate their artificial origin. \\
\begin{figure*}[ht!]
\centering
\includegraphics[width=\textwidth]{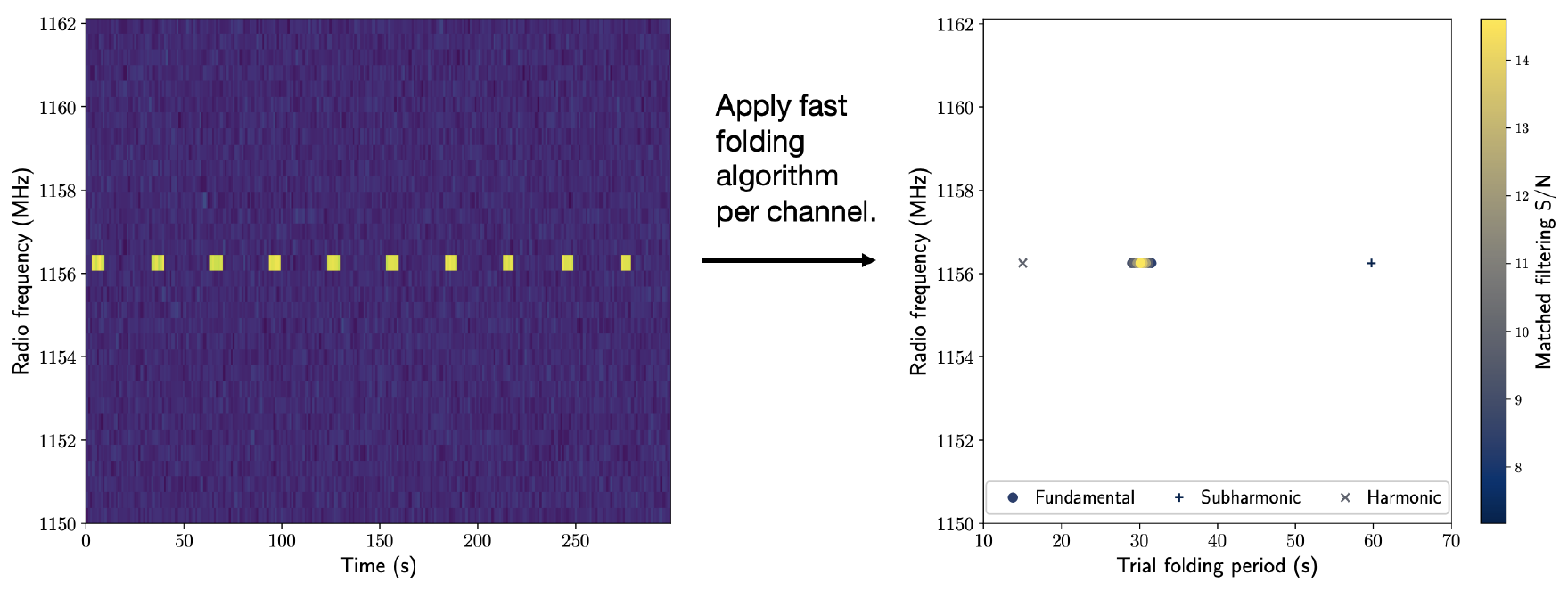}
\caption{Detection of a target periodic spectral signal using {\tt blipss}. Left: simulated radio frequency-time data containing 64 spectral channels of width 391~kHz each. A periodic pulsed signal has been inserted into the central channel. Right: radio frequency-period  ($P$) diagram showing the detection of a $P_0=30$~s signal in the central spectral channel. The first harmonic (cross) and the first subharmonic (plus) of the fundamental signal frequency ($f_0 = 1/P_0$) are also detected at $P=15$~s and $P=60$~s respectively. \label{fig1}}
\end{figure*}

On the Earth, periodic signals from human technology find extensive utility as radar for remote sensing, telecommunication, and aircraft navigation. Notable examples of air control radar include the Punta Salinas\footnote{\url{http://www.naic.edu/~phil/rfi/rdr/puntaSalinas/puntaSalinas.html}} and TPS-75\footnote{\url{http://www.mobileradar.org/radar_descptn_3.html\#tps_75}} radar operating in the United States between 1.3--1.7 and 2.9--3.1~GHz respectively. The former, with a radar rotation period of about 12~s, has often been picked up as periodic radio frequency interference (RFI) in historical transient searches (see Section~3.5 of  \citealt{Deneva2009}) with the 300-meter William E.~Gordon Arecibo radio telescope in Puerto Rico. \\

Though periodic signals are indispensable features of human technology, radio searches for periodic ETI transmissions have been scarce. \citet{Gray2002} looked for periodic transmissions in the locale of the ``Wow!'' signal, a strong narrowband emission detected near the 21~cm HI line in 1977 \citep{Kraus1979}. \citet{Harp2018} conducted an autocorrelation search for periodic extraterrestrial signals in 1--10~GHz data gathered with the Allen Telescope Array in California, USA. \\

Here, we present the Breakthrough Listen Investigation for Periodic Spectral Signals ({\tt blipss}), a novel software architecture utilizing a Fast Folding Algorithm (FFA: \citealt{Staelin1969,Morello2020}) to enable routine searches for periodic technosignatures in radio data. Our target signals are channel-wide periodic pulse trains, an example of which is illustrated in Figure~\ref{fig1}. Section~\ref{sec:energybudget} presents an energy budget comparison that evaluates the feasibility of ETI transmissions of our target signals against that of narrowband continuous-wave (CW) beacons and artificially dispersed broadband bursts. \\

Of all directions in the sky, the line of sight towards the Galactic Center (GC) offers the greatest propensity for the emergence of intelligent life \citep{Gajjar2021}. Further, radial beacons from the GC are viable given their advantageous positioning for galaxy-wide communications \citep{Benford2010}. Thus, the GC is an ideal first target for our periodic technosignature searches. Section~\ref{sec:obs} describes our 4--8~GHz observations undertaken as part of the Breakthrough Listen (BL) GC survey \citep{Gajjar2021}. We detail the methodology of {\tt blipss} in Section~\ref{sec:blipss}. In Section~\ref{sec:results}, we present the findings obtained from running {\tt blipss} on our GC data. Section~\ref{sec:discussion} highlights the implications of our study with emphasis on its potential outstanding gains for SETI. Finally, we summarize and conclude our investigation in Section~\ref{sec:conclusion}.

\section{Signal Transmission Energy Budgets} \label{sec:energybudget}
Consider an extraterrestrial antenna at the GC broadcasting three signal morphologies, i.e., steady Hz-wide CW beacons, broadband millisecond pulses, and channel-wide periodic pulse trains. For effective aperture $\eta_{\rm tx} = 0.7$ and telescope diameter $D_{\rm tx}=100$~m similar to Robert C.~Byrd Green Bank Telescope (GBT), the antenna gain is $G_{\rm tx} \approx \eta _{\rm tx}(\pi D_{\rm tx}c/\nu)^2 \approx 2.8 \times 10^7$ at $\nu = 6$~GHz. Here, $c$ is the vacuum speed of light.\\

Say that we receive extraterrestrial transmissions at the Earth using the GBT with system-equivalent flux density, $S_{\rm sys,GC} \approx 46.3$~Jy \citep{Suresh2022} towards the GC. For simplicity, let us assume that the transmitter and the receiver are perfectly aligned, have identical operating radio frequency bands, and have negligible signal digitization losses. \\

The minimum detectable EIRP of a narrowband CW beacon transmitted by the extraterrestrial antenna is \citep{Enriquez2017,Gajjar2022}
\begin{align}\label{eqn1}
{\rm EIRP}^{\rm min}_{\rm narrow} = & ({\rm S/N})_{\rm min}\beta S_{\rm sys,GC}\left( \frac{B_{\rm narrow}}{2T_{\rm int}}\right)^{1/2} \nonumber \\
& \times 4\pi d_{\rm GC}^2 . 
\end{align}
Here, $d_{\rm GC} \approx 8.18$~kpc is the distance to the GC \citep{GRAVITY2019}, and $B_{\rm narrow} \approx 1$~Hz is the CW signal bandwidth. The quantity $\beta$ is the signal dechirping efficiency. For receiver channel bandwidth $\Delta \nu_{\rm ch}$, sampling time $t_{\rm samp}$, and trial drift rate $\dot{\nu}$ \citep{Gajjar2021,Margot2021},
\begin{align}\label{eqn2}
\beta \approx \begin{cases}
1 &, ~ \dot{\nu} \leq \Delta \nu_{\rm ch}/t_{\rm samp} \\
\Delta \nu_{\rm ch}/ \dot{\nu} t_{\rm samp} &, ~ \dot{\nu} > \Delta \nu_{\rm ch}/t_{\rm samp}.
\end{cases}
\end{align}
Assuming perfect CW beacon detection, we take $\beta = 1$ throughout Section~\ref{sec:energybudget}. \\

Equation~\ref{eqn1} assumes a $\Delta \nu_{\rm ch}$ tuned to $B_{\rm narrow}$ for optimal detection. Using an integration time $T_{\rm int} = 5$~minutes and requiring a threshold signal-to-noise ratio $({\rm S/N})_{\rm min} =10$ for detection, ${\rm EIRP}^{\rm min}_{\rm narrow} \approx 1.5 \times 10^{17}$~W. Over a transmitter operation lifetime $L$, the net energy cost associated with transmission of a Hz-wide CW signal from the GC is
\begin{align}\label{eqn3}
W^{\rm min}_{\rm narrow} &= {\rm EIRP}^{\rm min}_{\rm narrow}L/G_{\rm tx} \nonumber \\
&\approx 2.7 \times 10^7~{\rm kW.h} \left(\frac{L}{\rm 300~s}\right). 
\end{align}

Similarly, for a single broadband burst \citep{Gajjar2022},
\begin{align}\label{eqn4}
{\rm EIRP}^{\rm min}_{\rm broad} = & ({\rm S/N})_{\rm min} S_{\rm sys,GC} \left( \frac{B_{\rm broad}}{2t_{\rm pulse}}\right)^{1/2} 4\pi d_{\rm GC}^2,
\end{align}
where $B_{\rm broad} = 3.42$~GHz is the detectable burst bandwidth limited by the usable radio frequency band of our observations (see Section~\ref{sec:obs}). Assuming a burst duration $t_{\rm pulse} = 1$~ms, we obtain ${\rm EIRP}^{\rm min}_{\rm broad} \approx 4.8 \times 10^{24}$~W for a single burst. For burst repetition with periodicity $P$, the net energy budget for signal transmission is
\begin{align}\label{eqn5}
W^{\rm min}_{\rm broad} &= {\rm EIRP}^{\rm min}_{\rm broad} \left\lceil \frac{L}{P} \right\rceil \frac{t_{\rm burst}}{G_{\rm tx}} \nonumber \\
&\approx 2.9 \times 10^{9}~{\rm kW.h} \left\lceil \frac{L}{P} \right\rceil \left( \frac{t_{\rm burst}}{1~{\rm ms}}\right). 
\end{align}
Here, $\lceil . \rceil$ denotes the ceiling function, and $\lceil L/P \rceil$ quantifies the number of pulses emitted in transmitter lifetime $L$. \\ 
\begin{figure}[t!]
\centering
\includegraphics[width=0.5\textwidth]{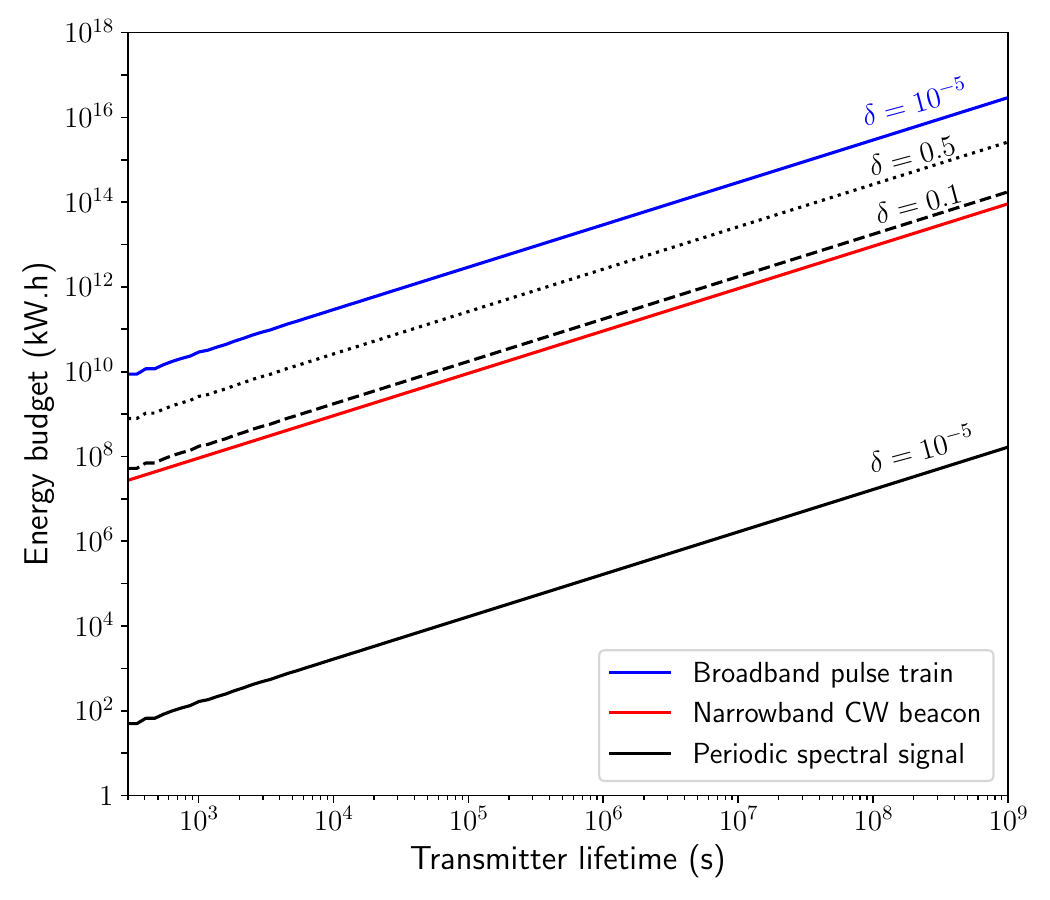}
\caption{Transmission energy budget comparison for ${\rm S/N}=10$ detection of three signal morphologies, i.e., steady Hz-wide CW beacons (red), a sequence of individually detectable broadband millisecond pulses (blue), and $\sim$ kHz-wide periodic spectral signals (black). The blue curve and all black curves assume a signal periodicity of 100~s. The blue and solid black curves both presume a pulse duty cycle, $\delta = 10^{-5}$. In contrast, the dashed and dotted black curves correspond to $\delta = 0.1$ and $0.5$ respectively. The red line assumes $100\%$ signal dechirping efficiency. Finally, the red line and all black curves reflect numbers for an integration time of 5~minutes. \label{fig2}}
\end{figure}

Finally, for an FFA detection of a channel-wide periodic signal with duty cycle $\delta = t_{\rm pulse}/P$ \citep{Morello2020}, 
\begin{align}\label{eqn6}
{\rm EIRP}^{\rm min}_{\rm periodic} = & ({\rm S/N})_{\rm min} \left(\frac{S_{\rm sys,GC}}{\mathcal{E}} \right) \left( \frac{\Delta \nu_{\rm ch}}{2T_{\rm int}}\right)^{1/2} \nonumber \\
& \times \left( \frac{\delta}{1-\delta}\right)^{1/2} 4\pi d_{\rm GC}^2.
\end{align}
Here, we consider $T_{\rm int} \geq 3P$ for periodicity detection. The quantities $ \mathcal{E}_{\rm FFA} \approx 0.93$ and $\Delta \nu_{\rm ch}\approx 2.86$~kHz are, respectively, the FFA search efficiency and our channel bandwidth (see Section~\ref{sec:obs}). Detection of periodic signals using a per-channel FFA requires that the channel bandwidth be matched to the signal bandwidth. \\

Transmission of kHz-wide periodic signals for detection with kHz-wide channels offers an advantageous arrangement for both the transmitter and the receiver. Where such a proposition allows the transmitter to avoid the elevated power costs of broadband signal transmission, the receiver benefits from negligible Doppler signal drift across channels. Assume that ETI correct for their own Doppler motion relative to the solar system barycenter. The largest contribution to any observed signal drift then comes from the Earth's daily rotation, which produces a maximum Doppler drift rate of $\approx 0.67$~Hz/s (see Section~3 of \citealt{Price2020}) at 6~GHz. Over $T_{\rm int}=5$~minutes, the resulting signal drift is $\approx 200~{\rm Hz} \lesssim \Delta \nu_{\rm ch} \sim$~kHz. \\

While our target signals are kHz-wide periodic pulse trains, a per-channel FFA is also sensitive to periodically modulated CW signals. For instance, consider a CW beacon turned on for 1~ms duration. The ensuing transmission has a sinc frequency response with a $\simeq 1$~kHz main-lobe bandwidth, i.e., comparable to our channel bandwidth. Periodic repetition of the modulation then produces a signal detectable via our per-channel FFA. \\

To facilitate consistent comparison between Equations~\ref{eqn4} and \ref{eqn6}, let us assume $\delta = 10^{-5}$, implying $t_{\rm pulse} = 1$~ms for $P=100$~s. Then, ${\rm EIRP}^{\rm min}_{\rm periodic} \approx 2.8 \times 10^{16}$~W. Applying Equation~\ref{eqn5} to channel-wide periodic signals, 
\begin{align}\label{eqn7}
W^{\rm min}_{\rm periodic}
&\approx 16.6~{\rm kW.h} \left\lceil \frac{L}{P} \right\rceil \left( \frac{t_{\rm pulse}}{1~{\rm ms}} \right).
\end{align}
Figure~\ref{fig2} compares transmission energy budgets at $L\geq 300$~s for ${\rm S/N} = 10$ detection of our three kinds of signals. Evidently, channel-wide periodic signals with $\delta = 10^{-5}$ offer the least energy cost for the transmitter. Hence, searches for periodic spectral signals from the GC present a promising avenue of exploration for SETI.

\section{Observations} \label{sec:obs}
\begin{figure}[t!]
\centering
\includegraphics[width=0.48\textwidth]{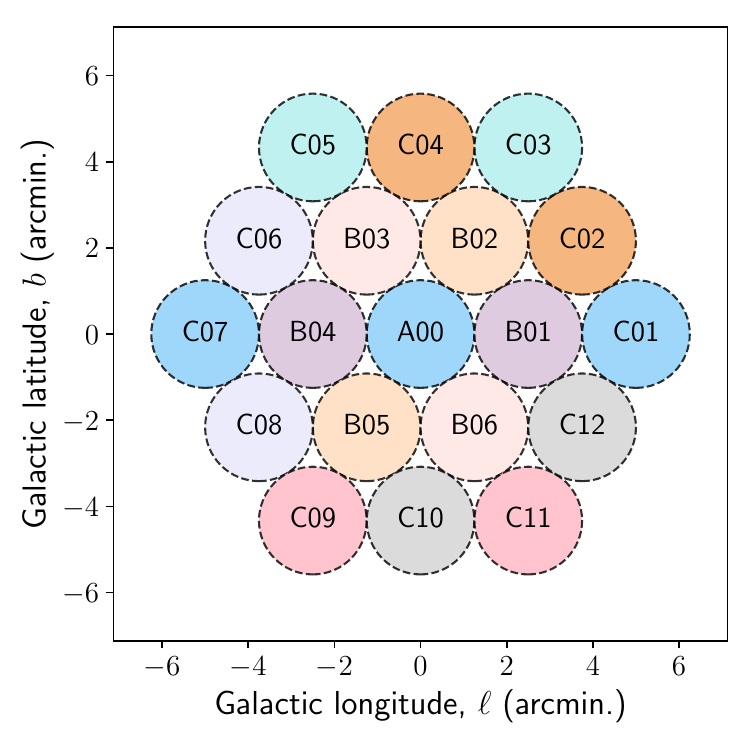}
\caption{4--8~GHz BL GC survey footprint covered by 19 spatial pointings (colored circles), each of HPBW $\approx 2\farcm5$. Various colors indicate pointings that have been grouped together for position-switched observations. \label{fig3}}
\end{figure}
\begin{deluxetable*}{CcccCcc}
\tablecaption{Log of 4--8~GHz GBT observations processed in our study. \label{tab1}}
\tablewidth{0pt}
\tablehead{
\colhead{Epoch} & \colhead{Calibrator}& \colhead{Test pulsar} & \colhead{Start date and time} & \colhead{Start MJD} & \colhead{Pointing Sequence\tablenotemark{a}} & \colhead{$N_{\rm scans}$\tablenotemark{b}} \\
\colhead{(number)} & \colhead{} & \colhead{} & \colhead{(UTC)} & \colhead{(UTC)} & \colhead{} & \colhead{}
}
\startdata
1 & 3C295 & B0355$+$54 & 2019 Aug 07 05:06:43.20 & 58702.2130 & (A00, C01, C07) & (3, 3, 2) \\
\hline
2 & 3C286 & B1133$+$16 & 2019 Aug 09 23:54:31.68 & 58704.9962 & (B01, B04) & (3, 3) \\
& & & 2019 Aug 10 00:25:29.28 & 58705.0177 & (B02, B05) & (3, 3) \\
& & & 2019 Aug 10 00:56:26.88 & 58705.0392 & (B03, B06) & (3, 3) \\
& & & 2019 Aug 10 01:27:24.48 & 58705.0607 & (C02, C04) & (3, 3) \\
& & & 2019 Aug 10 01:58:22.08 & 58705.0822 & (C03, C05) & (3, 3) \\
& & & 2019 Aug 10 02:29:19.68 & 58705.1037 & (C08, C06) & (3, 3) \\
& & & 2019 Aug 10 03:00:17.28 & 58705.1252 & (C11, C09) & (3, 3) \\
& & & 2019 Aug 10 03:31:32.16 & 58705.1469 & (C10, C12) & (2, 2) \\
\hline
\enddata 
\tablenotetext{a}{Observing sequence of pointings grouped for position switching}
\tablenotetext{b}{Number of scans per pointing. Each scan lasted 5~minutes.}
\end{deluxetable*}
\vspace{-1cm}
The BL GC survey, conducted by the Breakthrough Listen Initiative \citep{Worden2017,Gajjar2019}, is an extensive 0.7--93~GHz campaign of the GC and neighbouring Galactic bulge fields for technosignatures, pulsars, bursts, spectral lines, and masers \citep{Gajjar2021}. The 0.7--4~GHz component of the survey uses the CSIRO Parkes Murriyang radio telescope, while the 4--93~GHz portion utilizes the GBT. Here, we searched data from the completed 4--8~GHz segment of the survey for periodic spectral signals. \\

Figure~\ref{fig3} shows the 4--8~GHz BL GC survey footprint tiled by 19 distinct spatial pointings arranged in 3 concentric rings. From inner to outer, these rings are named A, B, and C respectively. All pointings made use of the single-beam C-band receiver, resulting in a half-power beam width (HPBW), $\theta_{\rm HPBW} \approx 2\farcm5$ at 6~GHz. Table~\ref{tab1} presents an overview of our observing program comprised of alternating scans between pairs or triplets of pointings with centers separated by at least $2\theta_{\rm HPBW}$. We also observed a test pulsar at every observing epoch to verify our system integrity. \\

All observations utilized the Breakthrough Listen Digital Backend \citep{MacMahon2018,Lebofsky2019} to channelize voltage data to filterbank products of various spectral and temporal resolutions. We refer readers to Table~4 of \citet{Gajjar2021} for details of data products generated by the BL GC survey. Here, for our target fields, we utilized the mid-spectral resolution data products having $\approx 1.073$~s time sampling. With a channel bandwidth of $\approx 2.86$~kHz, these data contain 1,703,936 channels covering 3.56--8.44~GHz, which overlaps the 3.9--8.0~GHz operating range of the C-band receiver. Our test pulsar data come in the mid-time resolution format with $\approx 349.53~\mu$s sampling time, $\approx 366.21$~kHz channel bandwidth, and 13,312~channels spanning 3.56--8.44~GHz. \\

Working with 4--8~GHz BL GC survey data, \citet{Gajjar2021} and \citet{Suresh2021,Suresh2022} reported the presence of significant radio frequency interference (RFI) between 4.24--4.39~GHz. Clipping bandpass edges and masking RFI-affected channels, we restricted our radio frequency coverage to 4.39--7.81~GHz, thus obtaining 3.42~GHz of usable bandwidth for our data analyses.

\section{{\tt blipss} Workflow} \label{sec:blipss}
The FFA is a phase-coherent time-domain search for periodic signals \citep{Staelin1969,Morello2020}. In comparison to traditional Fourier-domain periodicity searches involving incoherent harmonic summing, the FFA provides enhanced search sensitivity to short pulse duty cycles ($\delta \lesssim 0.1$). Combining the FFA with boxcar matched filtering of folded profiles, \citet{Morello2020} developed {\tt riptide}, a software package for real-time radio pulsar searches. Here, we adapted utilities from {\tt riptide} to build {\tt blipss}\footnote{Code available at \url{https://github.com/UCBerkeleySETI/blipss}.}, a CPU-based Python package to uncover periodic pulsed extraterrestrial signals in radio data. \\
\begin{figure*}[th!]
\centering
\includegraphics[width=\textwidth]{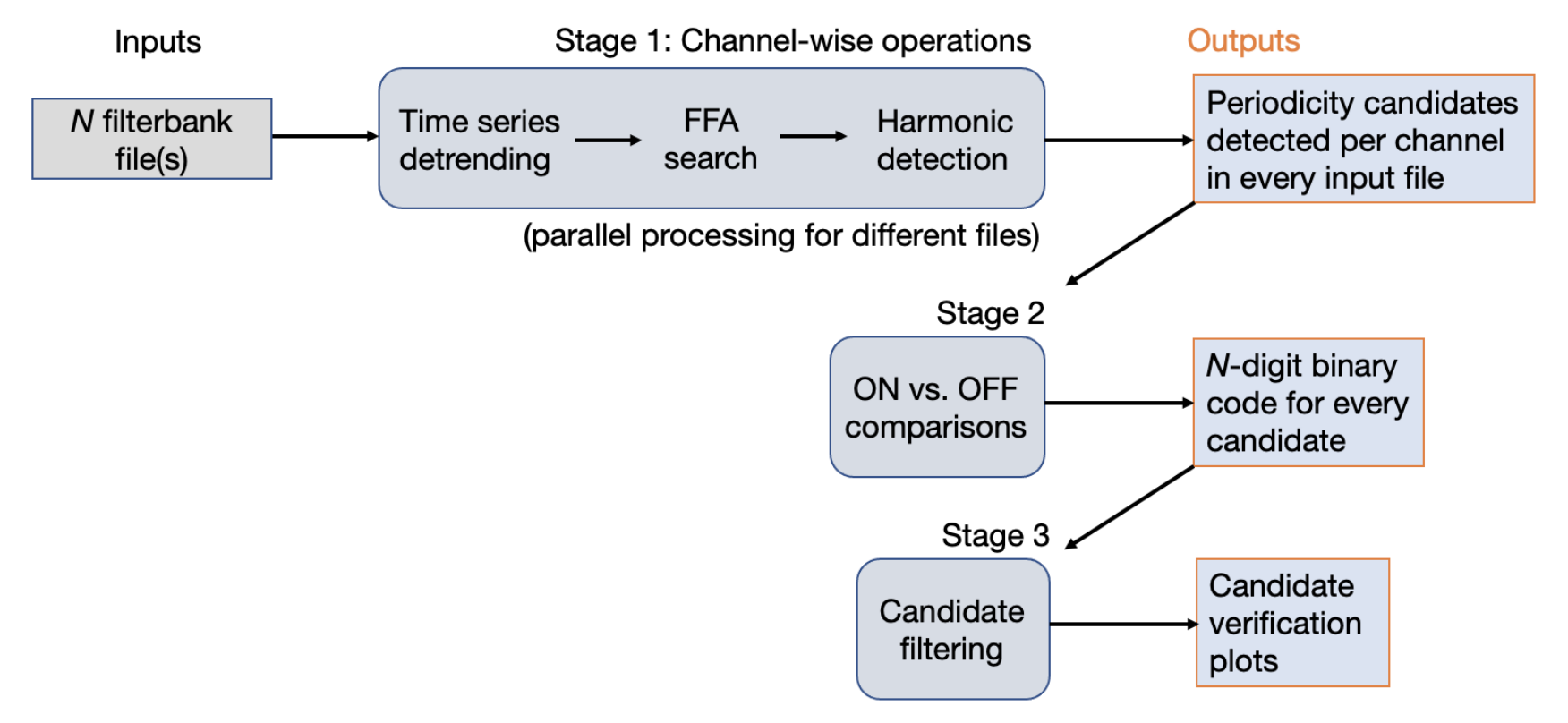}
\caption{A schematic representation of the {\tt blipss} workflow consisting of three stages. In Stage~1, a sequence of $N$ input filterbank files is processed channel-wise to search for periodic signals. Operations on different input files are parallelizable across multiple cores of a machine. Stage~1 outputs a list of periodicity candidates detected per channel for every input file. In Stage~2, these candidates are assigned $N$-digit binary codes based on comparisons between on-target and off-target scans. A ``1'' at entry~$j$ of the code (read from left to right) indicates candidate detection in the \mbox{$j$-th} input filterbank file. Similarly, we use zeroes to label candidate non-detections in different files. Finally, in Stage~3, we define filters to select binary codes warranting visual candidate inspection. \label{fig4}}
\end{figure*}

Figure~\ref{fig4} presents a flowchart of the {\tt blipss} workflow split into three stages. As input, our pipeline accepts one or more RFI-masked filterbank files. Dynamic spectra (radio frequency-time data) from these files are processed channel-wise to enable periodic signal detection regardless of signal bandwidth and dispersion. \\
\begin{figure*}[th!]
\centering
\includegraphics[width=\textwidth]{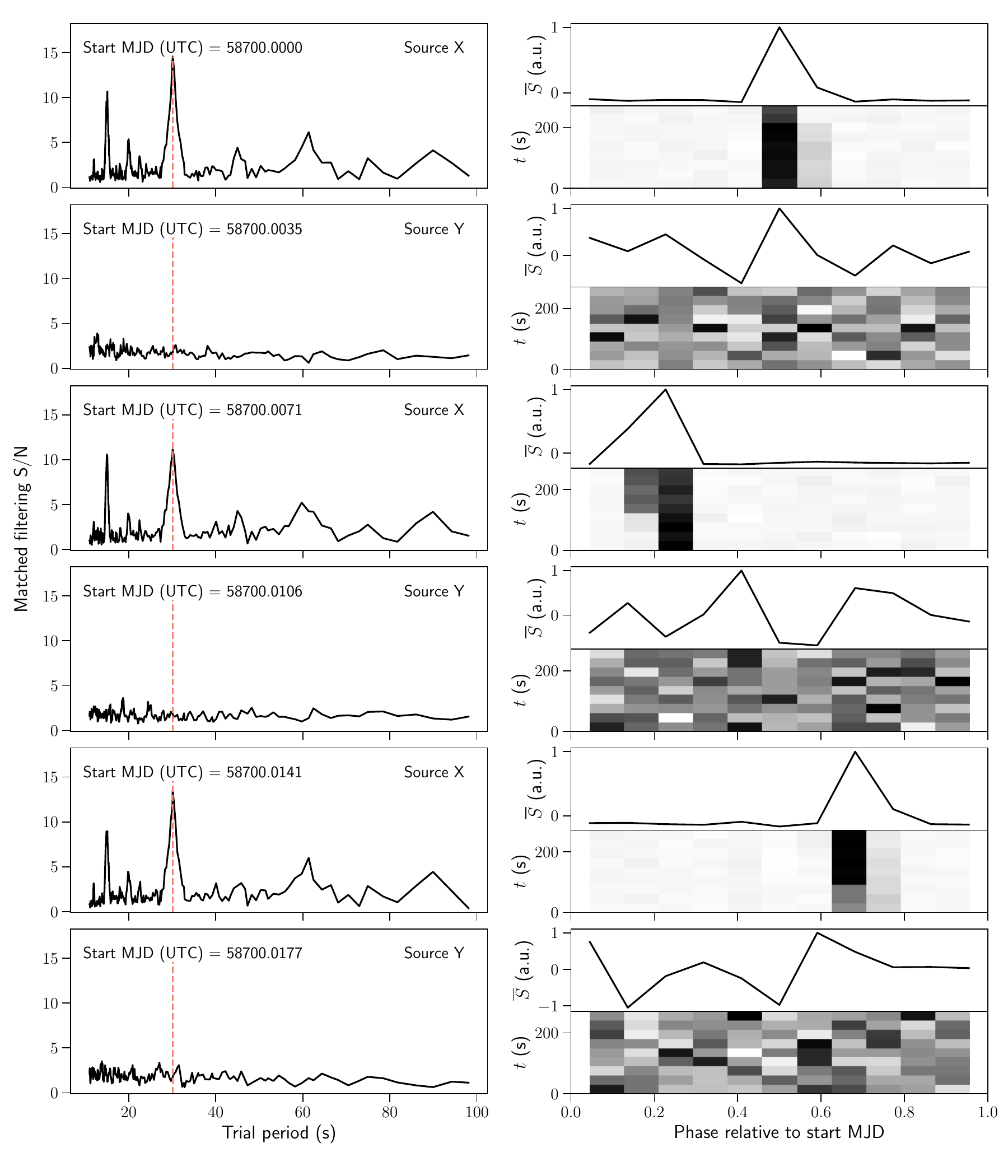}
\caption{A simulated candidate detected with 30~s period in all scans of source~X, but unseen in scans of source~Y. Left panels show periodograms generated from simulated single-channel time series data. Right panels illustrate phase-time diagrams (bottom subplots) and folded profiles (top subplots) of candidates at the periods highlighted by red vertical dashed lines in their respective left panels. Each row represents a different input data file. Rows $\{1, 3, 5\}$ and $\{2, 4, 6\}$ correspond to scans of sources X and Y respectively. \label{fig5}}
\end{figure*}
\begin{figure*}[t!]
\includegraphics[width=0.48\textwidth]{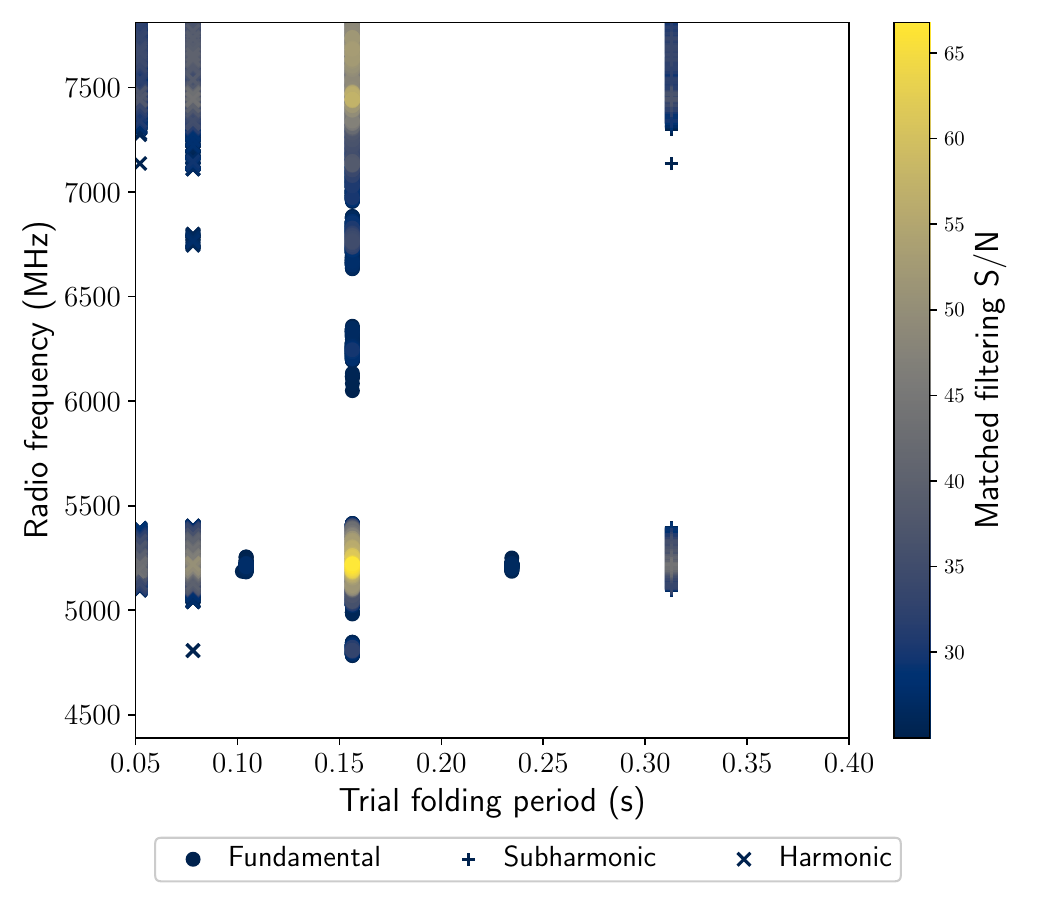} \qquad
\includegraphics[width=0.48\textwidth]{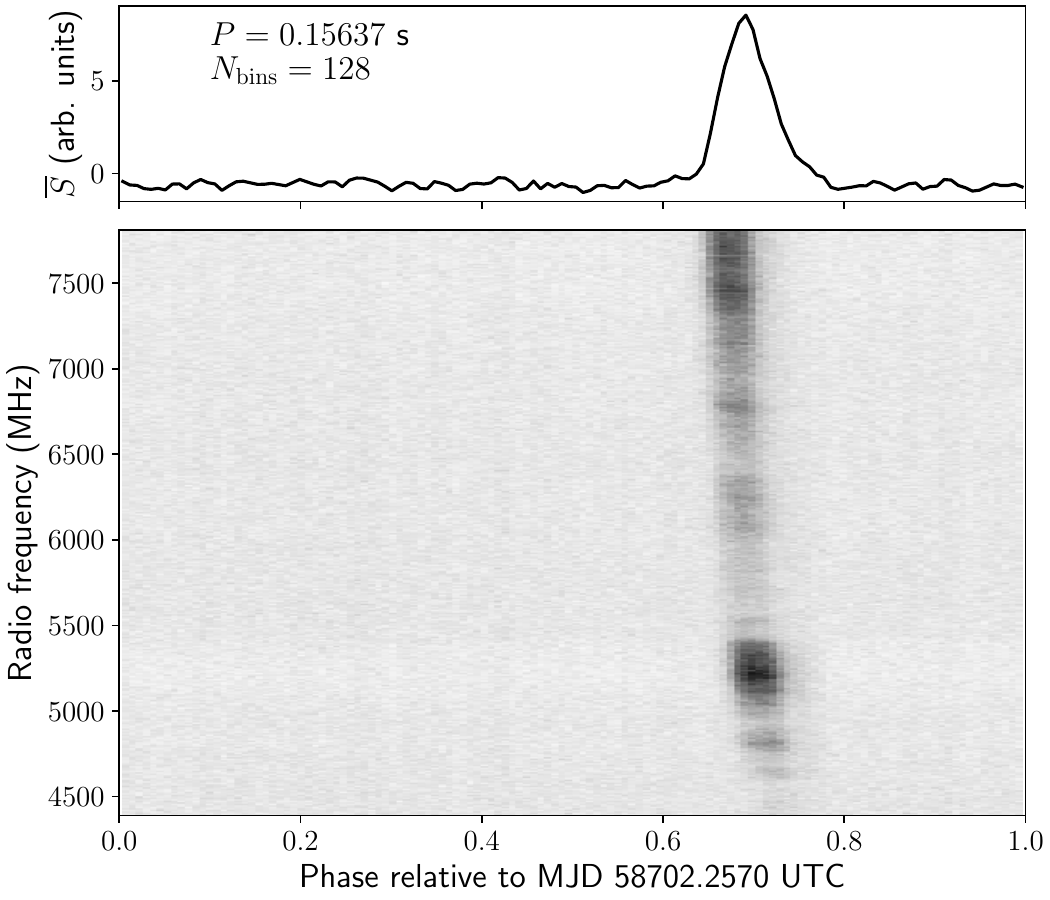}
\caption{Detection of test pulsar B0355$+$54 using {\tt blipss}. Left: Broadband signal detection at known pulsar rotational period, $P_0 \approx 156.37$~ms. We obtain our highest matched filtering S/N values (color scale) at radio frequencies where interstellar scintillation best brightens the pulsar emission. The first subharmonic ($P \approx 312.75$~ms) and the first two harmonics ($P \approx 78.19,~52.12$~ms) of the pulsar rotational frequency are also detected. In addition, we detect the pulsar at rational multiples of $P_0$ near 5.2~GHz, where the pulsar emission is at its brightest in our band. These include detections at $P \approx 2P_0/3$ and $P \approx 3P_0/2$. Right: Phase-resolved pulsar spectrum revealing pulse dispersion from radio wave propagation through the interstellar medium. Since our search strategy does not assume a dispersion law, {\tt blipss} can enable discovery of periodic signals with radio frequency drifts that differ from the often assumed cold plasma dispersion law (arrival times $ \propto \nu^{-2}$). \label{fig6}}
\end{figure*}

Consider a sequence of $N$ filterbank files generated via position switching between two hypothetical targets X and Y. Say that odd and even-numbered files contain data of sources X and Y respectively. In the first stage of our workflow, all data are detrended channel-wise using a running median filter to remove slow baseline fluctuations. This red noise removal step is essential to enhance our sensitivity to long periods. Next, the data per channel are searched for periodic signals using the FFA. For every channel in each input data set, periodicity candidates are then grouped into sets of harmonically related periods. Within every such set, the candidate period with the highest detection S/N is labeled as the fundamental. Harmonics and subharmonics of the fundamental frequency are then defined in the Fourier sense as follows.
\begin{align}
H_m &= P_0/(m+1), \label{eqn8} \\
S_n &= (n+1)P_0. \label{eqn9}
\end{align}
Here, $1/H_m$ and $1/S_n$ are, respectively, the \mbox{$m$-th} harmonic and the \mbox{$n$-th} subharmonic of the fundamental frequency, $f_0=1/P_0$. The final output from our first processing stage is a list of periodicity candidates detected per channel for every input file. Key user-supplied parameters to the FFA include the minimum and the maximum trial periods, the shortest and the largest trial duty cycles, and the chosen duty cycle resolution of the search. Given a range of trial periods and a sampling time, spacings between consecutive trial periods are determined using Equation~5 of \citet{Morello2020}. \\

In the second stage of our processing, we assign $N$-digit binary codes to periodicity candidates through comparisons between on-target and off-target scans. A ``1'' at the \mbox{$j$-th} element of a code (read from left to right) denotes candidate detection in the \mbox{$j$-th} input filterbank file. Similarly, zeroes indicate candidate non-detections in corresponding files. Say that $N =6 $ in our example of nodding observations between sources X and Y. A periodicity candidate with code 101010 would then be a SETI event of interest seen in all scans of target~X and undetected in scans of target~Y. \\

At our final processing step, we apply filters to shortlist candidates for visual inspection of their folded profiles. We define ``filter n'' ($F_n$) events as candidates that are detected in exactly $n$ scans of the target field, and and are unseen in all off-target scans. In our example with $N=6$, ON target~X, and OFF target~Y, candidates with binary code 101010 constitute the set of $F_3$ events. Figure~\ref{fig5} shows a sample $F_3$ candidate generated using simulated data and detected with {\tt blipss}. A $P=30$~s candidate is evident in all scans of source~X, but unseen in scans of source~Y.

\subsection{Test Pulsar Detection}\label{sec:testpsr}
We verified the integrity of our pipeline by confirming the detection of the two test pulsars observed in our survey. First, we detrended our data in every channel using a running median window of width 2~s. For a pulsar with rotational period $P_0$, we then searched for pulsations between $0.3P_0$ and $2.5P_0$. Further, mandating a minimum $N_{\rm bins} = 128$~phase bins across a folding period, we obtained a duty cycle resolution of $1/N_{\rm bins} \approx 0.78\%$. Allowing for boxcar widths extending up to 5 phase bins, we conducted matched filtering searches for folded profiles with duty cycles between $0.78\%$ and $3.91\%$. Setting a matched filtering S/N threshold of 25, we successfully detected both test pulsars B0355$+$54 and B1133$+$16 using {\tt blipss}. \\

Figure~\ref{fig6} shows our detection of pulsar B0355$+$54 from epoch~1 of observations. Notably, our algorithm is agnostic to any radio frequency drifts exhibited in the dynamic spectrum. Therefore, {\tt blipss} can enable discovery of periodic pulsed beacons with exotic swept frequency structures that are frequently missed by traditional broadband pulse search techniques assuming cold plasma dispersion (e.g., \citealt{Lazarus2015,Gajjar2021,Gajjar2022}).

\begin{deluxetable}{cc}
\tablecaption{Parameter values for {\tt blipss} searches of target fields \label{tab2}}
\tablewidth{0pt}
\tablehead{
\colhead{Parameter} &  \colhead{Value}
}
\startdata
Running median width, $W_{\rm med}$ & 12~s \\
Range of trial periods & 11--100~s \\
Pulse duty cycle resolution & 10\% \\
Range of trial duty cycles & 10--50\% \\
S/N threshold for ON pointings & 7 \\
S/N threshold for OFF pointings & 6 \\
\hline
\enddata
\end{deluxetable}
\vspace{-5mm}
\begin{deluxetable*}{ccCRRRR}
\tablecaption{Results from {\tt blipss} searches of target fields \label{tab3}}
\tablewidth{0pt}
\tablehead{
\colhead{ON pointing} &  \colhead{OFF pointing} & \colhead{$N_{\rm scans}$\tablenotemark{a}} & \colhead{$F_1$ count\tablenotemark{b}} & \colhead{$F_2$ count\tablenotemark{b}} & \colhead{$F_3$ count\tablenotemark{b}} & \colhead{$N_{\rm cands}$\tablenotemark{c}}\\
\colhead{} & \colhead{} & \colhead{(number)} & \colhead{(number)} & \colhead{(number)} & \colhead{(number)} & \colhead{(number)}
}
\startdata
A00 & C01 & 3 & 0 & 0 & 0 & 326\\
B01 & B04 & 3 & 9,005 & 0 & 0 & 5,044,721 \\
B02 & B05 & 3 & 205,681 & 0 & 0 & 4,110,738\\
B03 & B06 & 3 & 669,209 & 0 & 0 & 11,125,362\\
B04 & B01 & 3 & 181,664 & 1,213\tablenotemark{d} & 0 & 1,207,065\\
B05 & B02 & 3 & 39,921 & 0 & 0 & 5,067,100\\
B06 & B03 & 3 & 411,031 & 0 & 0 & 4,500,341\\
C01 & A00 & 3 & 1 & 0 & 0 & 426\\
C02 & C04 & 3 & 802 & 0 & 0 & 74,800\\
C03 & C05 & 3 & 176 & 0 & 0 & 2,560\\
C04 & C02 & 3 & 243 & 0 & 0 & 38,701\\
C05 & C03 & 3 & 14 & 0 & 0 & 9,305\\
C06 & C08 & 3 & 47 & 0 & 0 & 3,341\\
C07 & A00 & \phn 2\tablenotemark{e} & 0 & 0 & - & 227\\
C08 & C06 & 3 & 22 & 0 & 0 & 2,764\\
C09 & C11 & 3 & 75 & 0 & 0 & 1,207\\
C10 & C12 & 2 & 148 & 0 & - & 844\\
C11 & C11 & 3 & 14 & 0 & 0 & 5,527\\
C12 & C10 & 2 & 3 & 0 & - & 13,098\\
\hline
\enddata
\tablenotetext{a}{Number of ON-OFF scan pairs}
\tablenotetext{b}{$F_n$: Events detected in exactly $n$ ON scans and zero OFF scans, $n \in \{1, 2, 3\}$.}
\tablenotetext{c}{Overall number of detected candidates, including all possible binary codes.}
\tablenotetext{d}{All candidates detected at $P \approx 21.67$~s (see bottom left panel of Figure~\ref{fig7}).}
\tablenotetext{e}{Final two A00 scans were used as OFF for the two C07 scans.}
\end{deluxetable*}
\vspace{-1cm}
\subsection{{\tt blipss} Searches of Target Fields} \label{sec:target_searches}
Table~\ref{tab2} lists various parameter values chosen for our {\tt blipss} searches of paired target fields. Visually noticing the presence of significant power-law red noise at $f \lesssim 0.1$~Hz in power spectra of our channelized time series, we selected a running median filter of width, $W_{\rm med} =12$~s, for time series detrending. Desiring at least $N_{\rm bins} = 10$~bins across a folding period, we adopted a duty cycle resolution of $10\%$ for our FFA searches. Further, we set the shortest trial period of our searches at $P_{\rm min} = 11$~s to meet the requirement $P_{\rm min} \geq N_{\rm bins}t_{\rm samp}$, where $t_{\rm samp} \approx 1.073$~s is the sampling time of our data. Necessitating a minimum of three full periods within our data length, our scan duration of 5~minutes limits the maximum trial period of our FFA searches to $P_{\rm max} = 100$~s. Trialing boxcar filters spanning 1–5 phase bins, we explored pulse duty cycles, $\delta \in [10\%, 50\%]$ in our FFA searches. \\
\begin{figure*}[ht!]
\includegraphics[width=\textwidth]{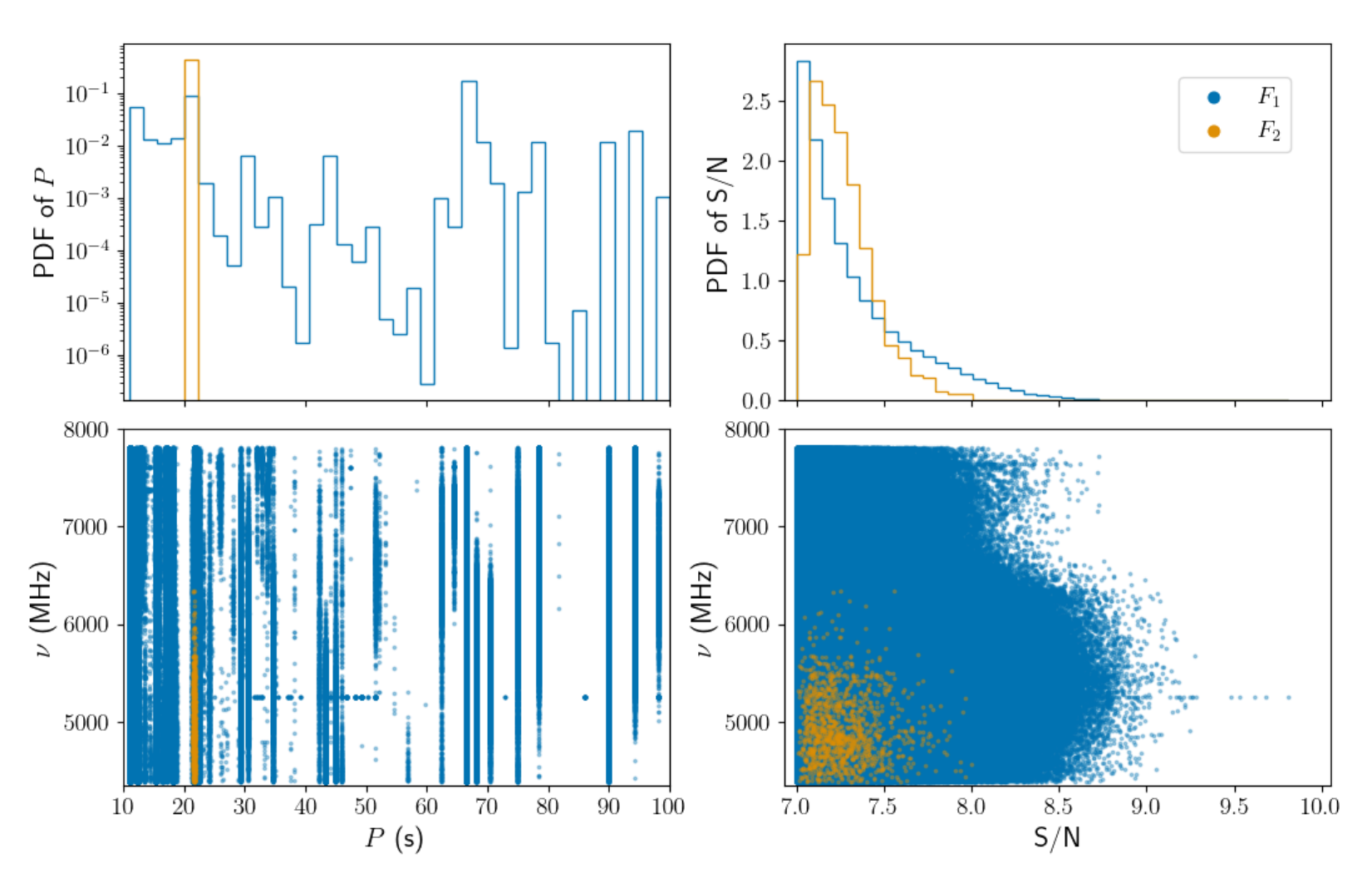}
\caption{Statistical distribution of $F_1$ (blue) and $F_2$ (orange) events detected in our survey. Top left panel: Probability density function (PDF) of candidate periods. The period axis from 11--100~s is spanned by 39~bins of width 2.28~s each. Top right panel: PDF of candidate S/N values. The S/N axis from 7--9.81 is covered by 39 equal-sized bins. Bottom left panel: Scatter of candidates in the radio frequency-period plane. Bottom right panel: Scatter of events in the radio frequency vs. matched filtering S/N diagram. \label{fig7}}
\end{figure*}

As indicated in Table~2, we adopted matched filtering S/N thresholds of 7 and 6, respectively, for our ON and OFF pointings. Here, motivated by \citet{Enriquez2017}, we chose a lower S/N threshold for our OFF scans to identify potential RFI candidates that may be weaker in our OFF observations. Such attenuation may occur for signals detected via telescope side lobes. Finally, we visually inspected folded profiles of all $F_n$ events, i.e., candidates found exclusively in ON pointings.
\begin{figure*}[th!]
\centering
\includegraphics[width=\textwidth]{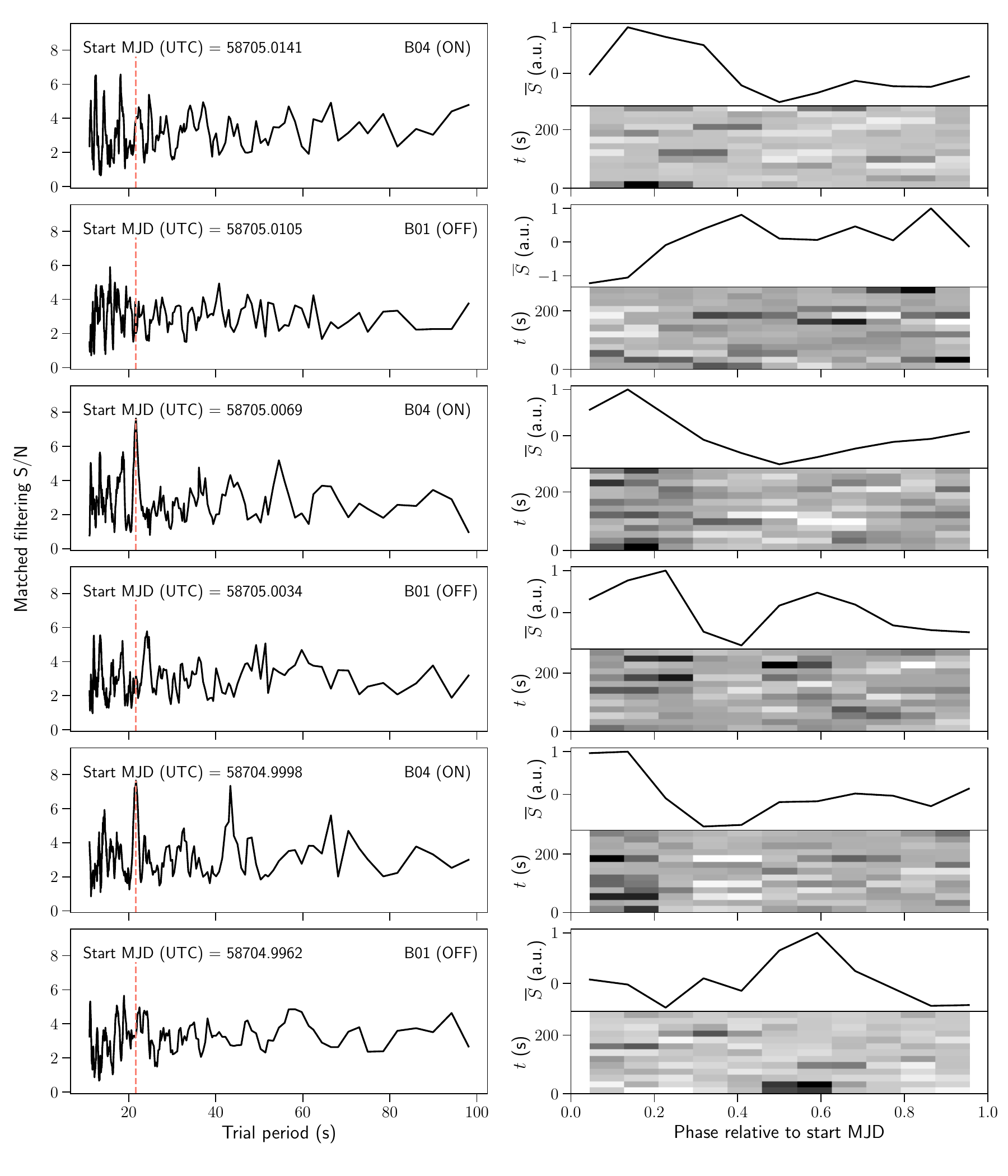}
\caption{An $F_2$ candidate with $P\approx 21.67$~s (vertical red dashed line in left panels) detected exclusively in two B04 scans (rows~3 and 5). Phase-time diagrams (bottom subplots of right panels) on rows~3 and 5 trace the candidate origin to the coincident alignment between intermittent RFI instances in our data. In particular, some of the aforementioned RFI are observed close to the start times of different observing scans. \label{fig8}}
\end{figure*}

\section{Results} \label{sec:results}
Table~\ref{tab3} summarizes the results of our FFA searches, yielding a total of 1,518,056~$F_1$ candidates and 1,213~$F_2$ events of interest across all pointings. Considering all possible binary codes, we notice significant variability in candidate counts across pointings. This finding likely suggests a dynamic RFI environment at the GBT. For instance, pointings A00, C01, and C07 that were jointly observed during epoch~1 (2019 Aug 07) have only about 200--400 candidates each. In contrast, pointings observed during epoch~2 have $\gtrsim$~1000~candidates each. Of note, all B-ring pointings with $\gtrsim 10^6$ candidates were observed in the first 1.5~hours of epoch~2. \\

In Figure~\ref{fig7}, we study the statistical distribution of $F_1$ and $F_2$ candidates in the $\nu$-$P$ and $\nu$-S/N diagrams. Evidently, all $F_2$ events were detected in B04~scans at $P\approx 21.67$~s. Figure~\ref{fig8} shows a sample $F_2$ candidate found in our study. Manually inspecting average pulse profiles of all such candidates, we attribute these events to coincident alignment between multiple instances of intermittent band-limited RFI. \\

Other notable features in Figure~\ref{fig7} include a narrowband line of events at 5.255~GHz and broadband streaks at multiple trial periods. Through visual candidate inspection, we identify the narrowband structure at 5.255~GHz with persistent RFI in our C02, C03, and C06 scans. We attribute the broadband features partly to noise and partly to bright transient RFI found near the edges of our observing scans. The observed accumulation of vertical striations towards the left edge of the \mbox{$\nu$-$P$} plane in Figure~\ref{fig7} is in part a statistical effect arising from the finer period spacing employed by the FFA at shorter trial periods. In summary, surveying the central $6\arcmin$ of our Galaxy, we report a non-detection of extraterrestrial spectral signals with $P \in [11~{\rm s},~100~{\rm s}]$  and $\delta \in [10\%,~50\%]$.

\section{Discussion} \label{sec:discussion}
\subsection{Number of Stars Surveyed}\label{sec:starcount}
The distribution of stars in the Milky Way constrains the number of potential transmitting worlds contained in the 4--8~GHz BL GC survey volume. Comparing different Galactic stellar density models, \citet{Gowanlock2011} argued that the \citet{Carroll2006} estimation closely reproduces the observed stellar density profile in the solar neighborhood. Let $r$ and $z$ denote, respectively, the radial distance from the GC and the vertical distance from the Galactic midplane. \citet{Carroll2006} modeled the stellar density in our Galaxy as
\begin{align}\label{eqn10}
n(r,z) = n_0e^{-r/h_r}(e^{-|z|/z_{\rm thin}} + 0.085e^{-|z|/z_{\rm thick}}). 
\end{align}
Here, $n_0 \approx 5.502$~stars~pc$^{-3}$ is a normalization constant \citep{Gowanlock2011}, and $h_r = 2.25$~kpc is the radial scale length. The constants $z_{\rm thin} = 350$~pc and  $z_{\rm  thick} = 1$~kpc are the vertical scale heights of the thin disk and the thick disk respectively. \\

At $d_{\rm  GC}=8.18$~kpc \citep{GRAVITY2019}, our $6\arcmin$ survey footprint radius equates to a projected extent, $\ell \approx 14.3$~pc. Since $\ell \ll z_{\rm thin}, z_{\rm thick}$, we can integrate $n(r,z)$ over $z$ assuming a constant Galactic midplane density between $z=\pm \ell$. The resulting radial stellar density profile is 
\begin{align}\label{eqn11}
n(r) \approx (170.5~{\rm stars~pc}^{-2})~e^{-r/h_r}.
\end{align}
Consider an observing cone from the Earth pointing towards the GC along the Galactic mid-plane. Using Equation~\ref{eqn11}, we estimated the number of stars in our survey cone as a function of distance from the Sun. Our 4--8~GHz BL GC survey volume contains nearly 8~million stars within a distance of 8.24~kpc. Of these, about 600,000 stars reside between distances of 8.12--8.24~kpc, i.e., at the GC.

\subsection{Search Sensitivity}\label{sec:sensitivity}
\begin{figure}[t!]
\centering
\includegraphics[width=0.5\textwidth]{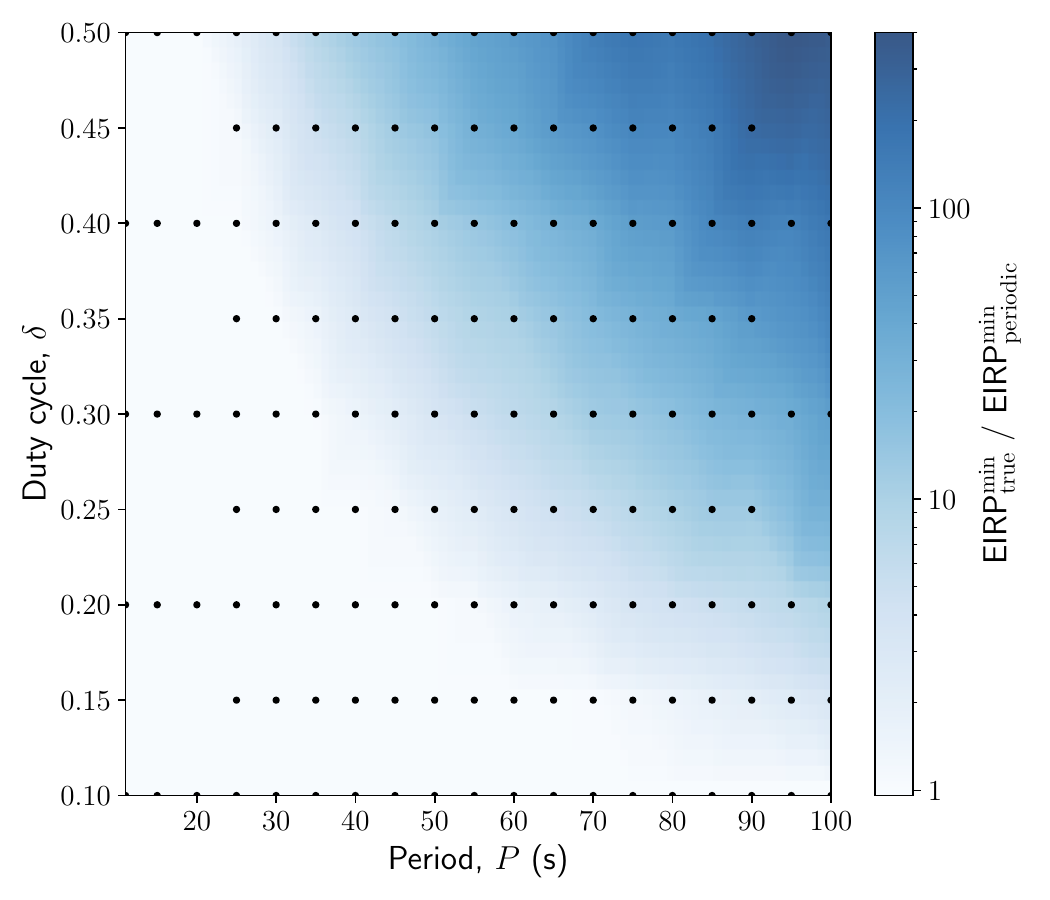}
\caption{Our true search sensitivity ${\rm EIRP}^{\rm min}_{\rm true}$ in the $P$-$\delta$ plane measured via injection of fake periodic signals into GBT time series data. The color scale quantifies ${\rm EIRP}^{\rm min}_{\rm true}$ relative to the theoretical sensitivity ${\rm EIRP}^{\rm min}_{\rm periodic}$ estimated from Equation~\ref{eqn6}. Black circular markers label trial ($P$,~$\delta$) of fake signal injections.\label{fig9}}
\end{figure}
Equation~\ref{eqn6} describes our theoretical survey sensitivity to periodic spectral signals, assuming an ideal Gaussian white noise background in time series data. However, real-world data frequently contain red noise that progressively hinders periodicity searches at large $\delta$ and long~$P$. Estimation of the true search sensitivity necessitates injection of fake periodic signals into real-world time series data and verification of their recovery. \\

Following Section~2.2 of \citet{Suresh2021}, we calibrated our A00 dynamic spectra to enable absolute survey sensitivity measurements. Injecting boxcar pulse trains into calibrated GBT time series data, we measured $7\sigma$ EIRP limits at several trial ($P$,~$\delta$). Performing 2D linear spline interpolation, Figure~\ref{fig9} illustrates our true search sensitivity ${\rm EIRP}^{\rm min}_{\rm true}$ in the $P$-$\delta$ plane. \\
\begin{figure}[t!]
\centering
\includegraphics[width=0.5\textwidth]{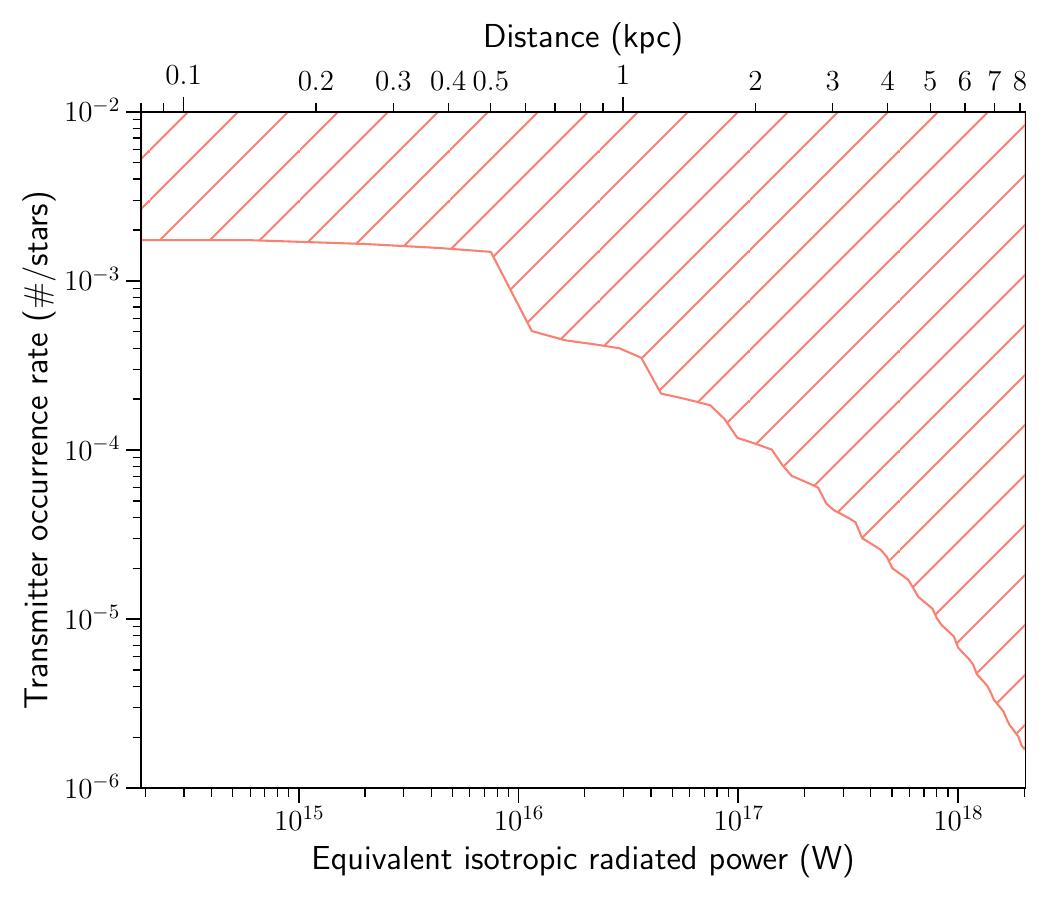}
\caption{Occurrence rate of 4--8~GHz extraterrestrial transmitters of kHz-wide periodic pulsed signals measured as a function of EIRP for a pulse duty cycle, $\delta \simeq 10\%$. The hatched region represents the parameter space excluded by our non-detection of channel-wide periodic signals in the 4--8~GHz BL GC survey. \label{fig10}}
\end{figure}

Noticeably, at $P \geq 60$~s and $\delta \geq 25\%$, the presence of red noise in our GBT data significantly elevates ${\rm EIRP}^{\rm min}_{\rm true}$ above the theoretical sensitivity ${\rm EIRP}^{\rm min}_{\rm periodic}$. Within our $P$-$\delta$ space in Figure~\ref{fig9}, ${\rm EIRP}^{\rm min}_{\rm true}$ reaches a maximum of  $\approx 2.5 \times 10^{21}$~W, i.e., $400 \times {\rm EIRP}^{\rm min}_{\rm periodic}$ at $P=100$~s and $\delta = 50\%$. Nevertheless, at $\delta \leq 15\%$, ${\rm EIRP}^{\rm min}_{\rm true} \approx {\rm EIRP}^{\rm min}_{\rm periodic}$ across all periods explored in our FFA searches. \\
\begin{figure*}[t!]
\centering
\includegraphics[width=\textwidth]{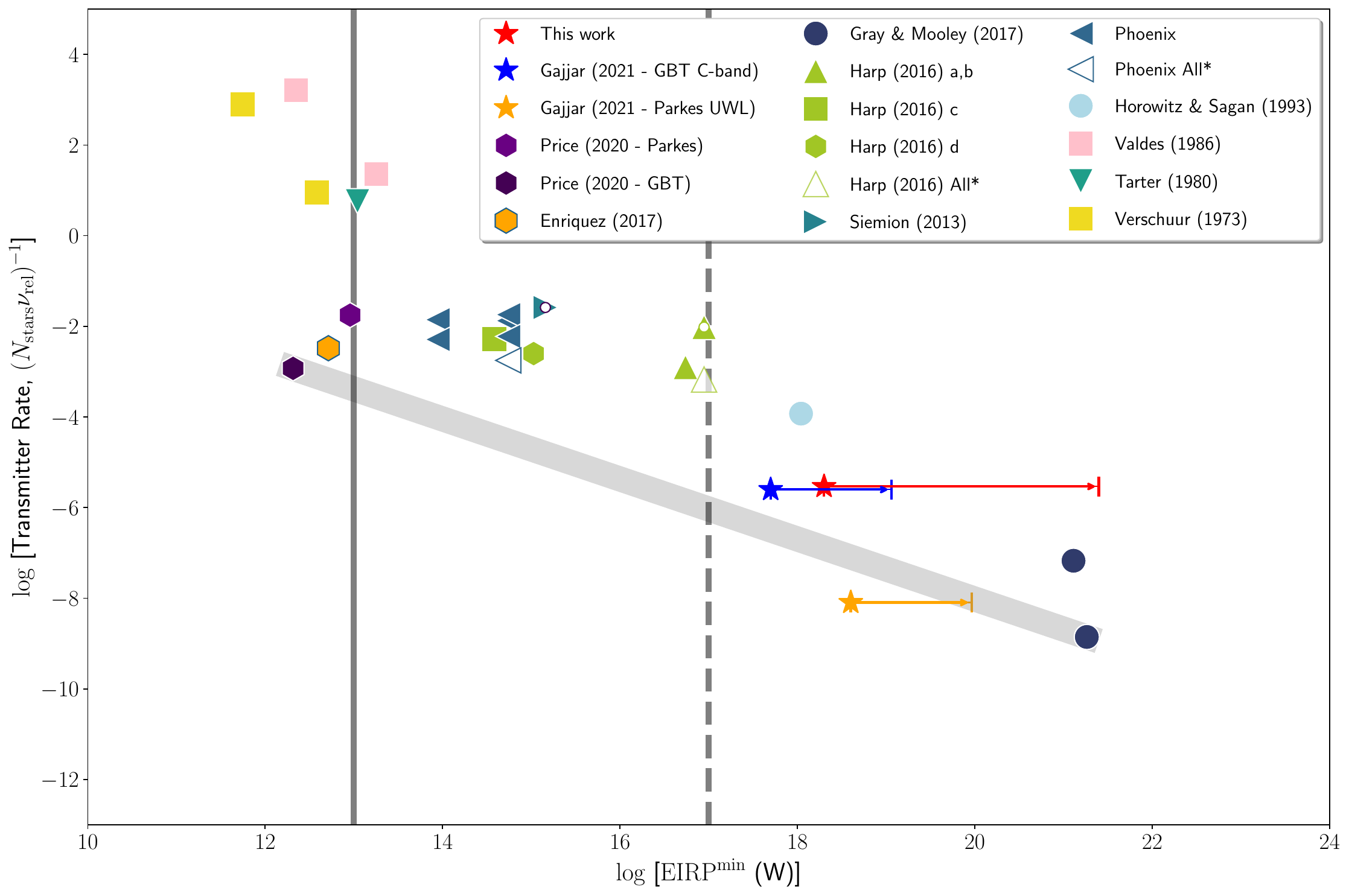}
\caption{Narrowband CW transmitter rate vs. minimum detectable EIRP for different surveys (various markers). Markers containing a white dot in their interior label surveys targeting exoplanets known to reside in the habitable zones around their host stars.  The solid gray vertical line marks ${\rm EIRP} \simeq 10^{13}$~W of an Arecibo-like transmitter. The dashed gray vertical line represents the  the total solar power incident on the Earth. The diagonal gray band is a power-law fit derived from the best constraining limits \citep{Gray2017,Price2020} on the CW transmitter rate at the low and high ${\rm EIRP}$ ends. Assuming a power-law distribution ($N_{\rm et} \propto {\rm EIRP}^{\alpha}$) of transmitters in our Galaxy, the slope $\alpha \approx -0.66$ of our power-law fit constrains the luminosity function of extraterrestrial CW transmitters between $\rm EIRP$ limits of $2 \times 10^{12}$~W to $2 \times 10^{21}$~W. Plotted points for past surveys are taken from Figure~13 of \citet{Gajjar2021}. \label{fig11}}
\end{figure*}

Figure~\ref{fig10} places limits on the occurrence rate of transmitters in our survey as a function of ${\rm EIRP}^{\rm min}_{\rm true}$ for $\delta \simeq 10\%$. Our investigations constrain the number of extraterrestrial worlds transmitting periodic spectral signals to fewer than one in about 600,000 stars at the GC above a minimum detectable EIRP of $\approx 2 \times 10^{18}$~W at $\delta \simeq 10\%$.  

\subsection{Comparisons with Past Surveys}\label{sec:surveys}
Our FFA searches bring novelty to radio SETI through their targeting of periodic spectral signals, a historically underexplored class of potential extraterrestrial transmissions. In contrast, traditional radio SETI programs have generally focused on the discovery of Doppler-drifting Hz-wide CW beacons \citep{Siemion2013,Harp2016,Tingay2016,Gray2017,Enriquez2017,Margot2018,Pinchuk2019,Price2020,Gajjar2021,Lacki2021,Margot2021,Traas2021,Franz2022,Garrett2023,Tusay2022} and artificially dispersed broadband bursts \citep{Gajjar2021,Gajjar2022}.  Consequently, an equal comparison of our search sensitivity against that of past radio SETI campaigns is difficult. However, our searches are sensitive to sufficiently bright periodically modulated CW signals, rendering CW transmitter rate comparisons against previous surveys viable. \\

A standard metric used in literature to evaluate the CW search performance of a survey is the Continuous Waveform Transmitter Figure-of-Merit (CWTFM) defined as
\citep{Enriquez2017}
\begin{align}\label{eqn12}
{\rm CWTFM} = \zeta_{\rm AO} \frac{\rm EIRP^{\rm min}}{N_{\rm stars}\nu_{\rm rel}}.  
\end{align}
Here, $\nu_{\rm rel}$ is the fractional bandwidth of observations, and ${\rm EIRP}^{\rm min}$ is the minimum EIRP detectable by the survey. For our FFA searches, ${\rm EIRP}^{\rm min}$ ranges between $\approx 2 \times 10^{18}$~W at $\delta = 10\%$ to $\approx 2.5 \times 10^{21}$~W at $(P,~\delta) = (100~{\rm s},~50\%)$. \\

The constant $\zeta_{\rm AO}$ is a normalization factor such that ${\rm CWTFM} = 1$ for an Arecibo-like transmitter with ${\rm EIRP} \simeq 10^{13}$~W, $\nu_{\rm rel} =0.5$, and $N_{\rm stars} = 10^3$. With $\nu_{\rm rel} \approx 0.57$ and $N_{\rm stars} \simeq 600,000$ at the GC, Equation~\ref{eqn12} implies CWTFM~$\simeq$~300--400,000 for our periodicity searches. In comparison, \citet{Gajjar2021} achieve CWTFM~$\simeq$~60--1400 in their dedicated narrowband signal searches of the GC at 4--8~GHz. \\

Lower CWTFM values correspond to more comprehensive searches for CW beacons. The product $(N_{\rm stars}\nu_{\rm rel})^{-1}$ is often termed the transmitter rate for narrowband CW beacons. Figure~\ref{fig11} compares the narrowband CW transmitter rate against ${\rm EIRP}^{\rm min}$ for various surveys. Past investigations marked in Figure~\ref{fig11} were dedicated searches for Doppler-drifting Hz-wide CW beacons. In contrast, our study, which targets channel-wide periodic signals, is sensitive to the periodicity at which potential CW beacons have been turned on or off. Therefore, while our search ${\rm EIRP}^{\rm min}$ overlaps with that of \citet{Gajjar2021}, Figure~\ref{fig11} presents an unequal comparison of our search methodology against past uniform narrowband CW beacon search strategies.

\subsection{Periodic Spectral Signal Transmitter Figure-of-merit}\label{sec:pssfom}
Motivated by CWTFM, we define a Periodic Spectral Signal Transmitter Figure-of-Merit (PSSTFM) to quantify our search completeness in relation to our processing parameters, our  survey volume, the telescope and the instrumentation used. For periodicity searches spanning $P \in [P_{\rm min},~P_{\rm max}]$ and  $\delta \in [\delta_{\rm min}, ~\delta_{\rm max}]$,
\begin{align}\label{eqn13}
{\rm PSSTFM} =~ &\frac{\xi_{\rm AO} \rm EIRP^{\rm min}}{N_{\rm stars}\nu_{\rm rel}\log{\left( \frac{P_{\rm max}}{P_{\rm min}} \right)} \log{\left( \frac{\delta_{\rm max}}{\delta_{\rm min}} \right)}}. 
\end{align}
Here, $\xi_{\rm AO}=10^{-10}$~W$^{-1}$ is a normalization constant such that ${\rm PSSTFM=1}$ when $\rm EIRP^{\rm min} = 10^{13}$~W, $N_{\rm stars}= 10^3$, $\nu_{\rm rel} = 0.5$, $P_{\rm max}/P_{\rm min} = 10^{2}$, and $\delta_{\rm max}/\delta_{\rm min} = 10$. For a given CWTFM, periodicity  searches that cover larger ranges in $P$ and $\delta$ have lower PSSTFM, and are therefore, said to be more complete. Our FFA searches achieve ${\rm PSSTFM} \sim 10^3$--$10^6$, where higher values reflect our increased $\rm EIRP^{\rm min}$ at longer $P$ and larger $\delta$.

\subsection{Future Work}\label{sec:futurework}
The coming decade promises to revolutionize radio SETI, with improved detection rates anticipated from the advent of ultra-wide bandwidth receivers, phased array feeds, and commensal backends for modern widefield sky surveys \citep{Houston2021}. The concomitant elevated data rates necessitate efficient automated pipelines for real-time event detection and follow-up. Enabling GPU acceleration of {\tt blipss} will be essential to permit large-scale real-time searches for periodic spectral signals in future SETI campaigns. \\

The Breakthrough Listen Initiative endeavors to continuously expand the search space of technosignatures explored in SETI. Where {\tt blipss} is sensitive to cyclic modulations of the total intensity, a natural extension to our SETI toolbox would be an autocorrelation search for signals that are cyclostationary in their complex voltages (see Section~2.10 of \citealt{Morrison2017}).

\section{Summary and Conclusion} \label{sec:conclusion}
Radio SETI has hitherto focused on the discovery of narrowband CW signals and artificially dispersed broadband pulses from extraterrestrial worlds. In contrast to these technosignature morphologies, periodic spectral signals offer an energetically efficient means of transmission across vast interstellar distances. A rotating beacon at the GC is, in particular, advantageously placed for galaxy-wide communications. \\

Here, we present {\tt blipss}, a CPU-based software package to enable FFA searches for periodic spectral signals from alien worlds. Operating on radio dynamic spectra, {\tt blipss} conducts FFA searches on a per-channel basis, thereby permitting periodicity detection regardless of signal bandwidth and dispersion. Consequently, {\tt blipss} can uncover signals with exotic swept frequency structures frequently missed by traditional broadband pulse search techniques that assume cold plasma dispersion. \\

Running {\tt blipss} on 4.5 hours of data from the 4--8~GHz BL GC survey, we report a non-detection of periodic pulsed signals with $P \in [11~{\rm s},~100~{\rm s}]$ and $\delta \in [10\%,~50\%]$. Thus, our investigations constrain the abundance of 4--8~GHz extraterrestrial transmitters of kHz-wide periodic spectral signals to fewer than one in about 600,000 stars at the GC above a $7\sigma$ EIRP threshold of $\approx 2 \times 10^{18}~$W at $\delta \simeq 10\%$. \\

Future developments to our work include GPU acceleration of routines in {\tt blipss} and incorporation of coherent voltage folding to expand our search space of periodic signals. Progress on the former will enable integration of {\tt blipss} into real-time search pipelines for large-scale event discovery and follow-up.  

\begin{acknowledgments}
Breakthrough Listen is managed by the Breakthrough Initiatives, sponsored by the Breakthrough Prize Foundation. P.N. was funded as a participant in the Berkeley SETI Research Center Research Experience for Undergraduates Site, supported by the National Science Foundation under Grant No. 1950897. S.Z.S. acknowledges that this material is based upon work supported by the National Science Foundation MPS-Ascend Postdoctoral Research Fellowship under Grant No.~2138147. The Green Bank Observatory is a facility of the National Science Foundation, operated under cooperative agreement by Associated Universities, Inc.
\end{acknowledgments}

\facility{GBT}

\software{Astropy \citep{Astropy2013, Astropy2018},
          Blimpy \citep{Blimpy},
          {\tt blipss} (this work),
          Matplotlib \citep{Matplotlib},
          NumPy \citep{NumPy},
          Python~3 (\url{https://www.python.org}),
          Pandas \citep{pandas2010, pandas2022},
          {\tt riptide-ffa} \citep{Morello2020},
          SciPy \citep{SciPy}.
         }
         


\bibliography{references}{}
\bibliographystyle{aasjournal}



\end{document}